# Gene set proximity analysis: expanding gene set enrichment analysis through learned geometric embeddings

Henry Cousins[1,2,*], Taryn Hall[3], Yinglong Guo[3], Luke Tso[3], Kathy Tzy-Hwa Tzeng[3], Le Cong[4,5], and Russ Altman[1,4,6,7,*]

**Affiliations:**
[1]Department of Biomedical Data Science, Stanford University School of Medicine, Stanford, CA, USA
[2]Medical Scientist Training Program, Stanford University School of Medicine, Stanford, CA, USA
[3]Optum Labs at UnitedHealth Group, Minneapolis, MN, USA
[4]Department of Genetics, Stanford University School of Medicine, Stanford, CA, USA
[5]Department of Pathology, Stanford University School of Medicine, Stanford, CA, USA
[6]Department of Medicine, Stanford University School of Medicine, Stanford, CA, USA
[7]Department of Bioengineering, Stanford University, Stanford, CA, USA

*Correspondence to: cousinsh@stanford.edu (H.C.); russ.altman@stanford.edu (R.A.)

## ABSTRACT

Gene set analysis methods rely on knowledge-based representations of genetic interactions in the form of both gene set collections and protein-protein interaction (PPI) networks. Explicit representations of genetic interactions often fail to capture complex interdependencies among genes, limiting the analytic power of such methods. Here we propose an extension of gene set enrichment analysis to a latent feature space reflecting PPI network topology, called gene set proximity analysis (GSPA). Compared with existing methods, GSPA provides improved ability to identify disease-associated pathways in disease-matched gene expression datasets, while improving reproducibility of enrichment statistics for similar gene sets. GSPA is statistically straightforward, reducing to classical gene set enrichment through a single user-defined parameter. We apply our method to identify novel drug associations with SARS-CoV-2 viral entry. Finally, we validate our drug association predictions through retrospective clinical analysis of claims data from 8 million patients, supporting a role for gabapentin as a risk factor and metformin as a protective factor for COVID-19 hospitalization.

**INTRODUCTION**

High-throughput sequencing and genetic perturbation methods produce vast functional genetic datasets representing both clinical and molecular phenotypes, with applications from drug discovery to clinical diagnostics (Cieślik and Chinnaiyan, 2017). While gene-level expression data is useful for identifying differentially expressed genes or training predictive models, assigning biological relevance to observed expression patterns generally requires inference of perturbation in higher-order functional pathways (De Leeuw *et al.*, 2016). This requires both defining the composition of such pathways and recognizing subtle expression signatures in noisy data.

Methods for analyzing expression changes in predefined gene sets generally fall into one of two algorithmic classes. These include overrepresentation approaches, which detect pathway enrichment within a set of differentially expressed genes, and aggregate-score approaches, which associate input genes with a continuous phenotypic score. Overrepresentation approaches, including ORA, DAVID, and Enrichr, benefit from simpler input requirements and faster runtimes due to compatibility with deterministic significance tests (Boyle *et al.*, 2004; Huang *et al.*, 2009; Chen *et al.*, 2013). However, the requirement for a binary threshold for differential expression limits their ability to detect pathway changes resulting from weaker differential expression among many genes, as well as the relative ordering of genes.

Aggregate-score approaches, in contrast, weight gene sets according to a cumulative statistic based on continuous phenotypic scores for genes. The most popular aggregate-score approach is gene set enrichment analysis (GSEA), which computes a weighted Kolmogorov-Smirnov statistic representing enrichment of a gene set in an ordered gene list (Subramanian *et al.*, 2005). This enrichment statistic is then compared to a null distribution generated from random permutations of the gene scores to assign a significance level. Aggregate-score approaches such as GSEA, CAMERA, and PADOG, are more sensitive to large ensembles of genes acting in concert but only consider overlap between explicitly defined genes in a set and input list (Wu and Smyth, 2012; Tarca *et al.*, 2012). As a result, such methods cannot detect pathway perturbations implied from expression changes in neighboring genes, which is particularly important in the case of noisy datasets or incomplete gene sets (Goeman and Bühlmann, 2007).

To overcome these limitations, some gene set analysis methods incorporate a priori assumptions about genetic relationships, typically in the form of protein-protein interaction (PPI) networks (Szklarczyk *et al.*, 2017; Hillenmeyer *et al.*, 2016). This enables gene-level enrichment statistics to be augmented by expression measures of a local genetic neighborhood. Due to the computational expense of repeatedly traversing large graphs such as PPI networks, nearly all network-based methods use the overrepresentation approach (Miryala *et al.*, 2018). Furthermore, such methods generally utilize explicit representations of network topology, such as one-hop neighbors of differentially expressed genes, which are limited in their ability to capture latent local and global network features. As a result, gene set analysis methods remain sensitive both to experimental noise and to the specific composition of gene sets under consideration, even among gene sets denoting the same pathway.

We hypothesize that capturing the full extent of functional pathway enrichment in an expression dataset requires consideration of complex pathway structures that are not efficiently represented in explicit form. Therefore, we propose a generalization of the classical GSEA algorithm to a latent feature space derived from unsupervised embeddings of a PPI network. Our method, called gene set proximity analysis (GSPA), implicitly considers the full network context of individual genes, allowing detection of perturbations in bona fide functional pathways even in noisy or incomplete datasets. Notably, GSPA is statistically straightforward, reducing precisely to classical GSEA through a single parameter. We find that GSPA outperforms both GSEA and NGSEA, a state-of-the-art network-augmented gene set analysis method, in identifying disease-associated pathways from gold-standard expression datasets, while also improving reproducibility for semantically similar gene sets.

Finally, we apply GSPA to a collection of datasets measuring gene involvement in SARS-CoV-2 viral entry using gene sets representing known targets of FDA-approved drugs, identifying four drugs (gabapentin, metformin, lorazepam, and clonazepam) as likely modulators of SARS-CoV-2 viral entry. We subsequently investigate our predictions through propensity-score matched, retrospective analysis of health insurance claims from 8 million patients. Consistent with the results from GSPA, our clinical investigation supports a role for gabapentin and metformin as risk and protective factors for severe SARS-CoV-2 infection, respectively, providing insight into both molecular pathogenesis and potential treatment strategies for COVID-19.

**METHODS**

*Generating gene embeddings*

To generate representative embeddings for protein-coding genes, we obtained high-confidence protein-protein interactions from the STRING repository (version 11; Szklarczyk *et al.*, 2017). We included only interactions with a confidence score of at least 0.9 in humans, resulting in a network of 12,396 proteins and 324,152 interactions. This was represented as an undirected graph, $G = (V, E)$, with $V$, the set of vertices, representing unique proteins and $E$, the set of edges, representing pairwise PPIs. Low-dimensional feature learning was performed on $G$ using the node2vec algorithm, which learns node embeddings by simulating biased random walks in $G$ to preserve local neighborhood architecture (Grover and Leskovec, 2016).

*Calculation of enrichment score*

For a given gene set and ranked gene list, the enrichment score in GSPA is computed in a similar manner as in GSEA, with an important modification. In GSEA, a weighted running sum statistic is computed by walking down the ranked list, incrementing the statistic in proportion to its phenotypic score when encountering a gene in the gene set and decrementing it otherwise. The ES is then the signed maximum absolute value of the statistic. Equivalently,

$$ES_k^{GSEA} = \sup_{1 \leq i \leq n} \left( \frac{\sum_{t=1}^{i} s_t \cdot \mathbf{1}_{g_t \in G_k}}{\sum_{t=1}^{n} s_t \cdot \mathbf{1}_{g_t \in G_k}} - \frac{\sum_{t=1}^{i} \mathbf{1}_{g_t \notin G_k}}{n - |G_k|} \right)$$

where $ES_k^{GSEA}$ is the enrichment score for gene set $G_k$, *sup* is the supremum, $|G_k|$ is the number of genes in the gene set, $i$ represents a position in the ranked gene list, $n$ is the number of genes in the gene list, $s_t$ is a normalized phenotypic score for a given gene $g_t$, and **1** is the indicator function for membership of $g_t$ with respect to $G_k$.

In GSPA, while calculating the running sum statistic, the gene set $G_k$ is temporarily augmented to create the set of gene-set-proximal genes $P_k$ such that

$$P_k = \{g \in L : \min(dist(v_g, V_{G_k})) \leq r\}$$

where $g$ represents any gene in the ranked list $L$, $v_g$ represents the embedding for $g$, $V_{Gk}$ represents a list of embeddings for each gene in $G_k$, *dist* outputs a list of cosine distances from $v_g$ to each embedding in $V_{Gk}$, and $r$ is a user-defined parameter representing the radius from which to expand each member of the original gene set. The raw ES is then computed as in Equation (1), substituting $P_k$ for $G_k$.

$$ES_k^{GSPA} = \sup_{1 \leq i \leq n} \left( \frac{\sum_{t=1}^{i} s_t \cdot \mathbf{1}_{g_t \in P_k}}{\sum_{t=1}^{n} s_t \cdot \mathbf{1}_{g_t \in P_k}} - \frac{\sum_{t=1}^{i} \mathbf{1}_{g_t \notin P_k}}{n - |P_k|} \right)$$

Importantly, this reduces to Equation (1), the definition of ES in GSEA, as $r$ decreases to zero.

*Generating null distributions*

The generation of a null distribution of enrichment scores for a given gene set is important both for assigning relative rankings to gene sets of different sizes and for assigning significance levels. In the GSEA algorithm for pre-ranked gene lists, null distributions are generated by sampling a random gene set $G_k$' containing the same number of members in the ranked list as the original set $G_k$ and recalculating ES. This implicitly defines a null hypothesis of no association between genes, which, for large gene sets, can result in highly sensitive estimates of significance at the expense of specificity. Therefore, by default, GSPA generates null distributions by first resampling the original gene set to create $G_k$', then creating a null set of proximal genes $P_k$' as in the original ES calculation for GSPA. A null ES is defined from $P_k$', and this procedure is repeated a fixed number of times. Alternatively, users can test a less stringent null hypothesis by directly resampling $P_k$ itself. Both methods reduce precisely to the original GSEA prerank algorithm as $r$ decreases to zero, but the former method directly accounts for known correlations between genes. Once the ES and null ES distribution have been calculated, normalized enrichment score, p-value, and false discovery rate are calculated as in GSEA.

*Datasets*

To evaluate the performance of GSPA, we used the gold-standard GEO2KEGG compendium, composed of 42 disease-matched, human gene expression profiles obtained from Bioconductor (Tarca *et al.*, 2012, 2013). Each dataset contains microarray results from an AffyMetrix HG-U133a chip and is mapped to one of 19 different diseases. Furthermore, each disease is accompanied by a predefined set of disease-associated KEGG pathways obtained from the MalaCards database of disease-gene associations. These datasets have been used in previous comparisons of gene set analysis methods (Tarca *et al.*, 2013; Geistlinger *et al.*, 2021, 2016).

Pathway gene sets were obtained from the human KEGG database (Kanehisa *et al.*, 2017). Gene sets containing fewer than 3 or greater than 600 genes were excluded from all analyses, resulting in a total of 332 KEGG gene sets considered.

*Running GSEA and NGSEA*

We used the Python implementation of GSEA, GSEApy (version 0.10.5), which is available at https://github.com/zqfang/GSEApy. Specifically, we used the weighted GSEA prerank method with the default parameters, as accepting a list of continuously scored genes as an input accommodates a wider range of experimental methods than does a gene expression matrix. We ran NGSEA with the default parameters using the NGSEA web server (https://www.inetbio.org/ngsea/index.php).

*Benchmarking prediction of disease-associated gene sets*

To measure the ability of GSPA to identify disease-associated pathways, we used a version of the procedure proposed by Geistlinger et al. (2021) for evaluating the biological relevance of gene set analysis methods. For each disease represented in our collection of evaluation datasets, we established a set of "ground-truth" KEGG pathways known to exhibit a strong association with the disease (specifically, a relevance ranking > 20 in the GSEABenchmarkeR package). We subsequently ran GSPA, GSEA, and NGSEA on each expression dataset in the GEO2KEGG compendium using KEGG pathway gene sets. For each method, KEGG pathways were ranked by normalized enrichment score, as in Geistlinger et al. (2016), which better accounts for differences in gene set size than does the raw enrichment score. We then measured the ability of each method to retrieve "ground-truth" gene sets using area under the precision-recall curve. Significance levels for each method were calculated using Wilcoxon signed-rank tests.

*Benchmarking reproducibility of gene set rankings*

To measure the extent to which similar rankings are preserved for gene sets representing the same pathway, we first defined a set of 54 unique, semantically similar gene sets. Specifically, we used a fully automated string matching procedure comparing KEGG gene set names by token set ratio after removing ID number and common terms such as "signaling" or "metabolism". For instance, "hsa00100_Steroid_biosynthesis" was matched with "hsa00140_Steroid_hormone_biosynthesis," and "hsa04136_Autophagy" was matched with "hsa04140_Autophagy." We then computed gene set rankings based on NES for GSPA, GSEA, and NGSEA and compared the similarity between predictions for matched gene sets by Spearman rank correlation coefficient.

*Predicting pharmacologic modulators of SARS-CoV-2 entry*

To assess the suitability of GSPA for predicting relevant drugs from gene expression data, we obtained disease-drug associations from the Comparative Toxicogenomics Database for each disease represented in the evaluation datasets (Davis *et al.*, 2017). We obtained drug-target gene sets from the DSigDB D1 database and measured the ability of GSPA gene-set rankings to retrieve known disease-drug associations by area under precision recall curve (Yoo *et al.*, 2015).

We obtained ranked gene lists from three genome-wide CRISPR knock-out screens of host gene importance for SARS-CoV-2 viral entry (Daniloski *et al.*, 2021; Schneider *et al.*, 2021; Wei *et al.*, 2021). We performed GSPA separately on each dataset using DSigDB D1 gene sets.

*Claims database*

The study sample was obtained from de-identified administrative claims for Medicare Advantage Part D (MAPD) enrollees in a research database. The database contains medical (emergency, inpatient, and outpatient) and pharmacy claims for services submitted for third party reimbursement, available as International Classification of Diseases, Tenth Revision, Clinical Modification (ICD-10-CM), and National Drug Codes (NDC) claims, respectively. These claims were aggregated after completion of care encounters and submission of claims for reimbursement.

*Cohort construction*

For each drug of interest, we constructed a cohort of individuals with at least 11 months of enrollment in MAPD insurance from January through December 2019 and at least 1 month of enrollment in MAPD in 2020. These individuals had at least one pharmacy prescription claim during their enrollment and lived in counties in New York, New Jersey, and Connecticut. In our database, COVID-19 hospitalization was more prevalent among individuals insured through MAPD and among residents of the New York, New Jersey, and Connecticut tri-state area. We restricted our analyses to these populations to select for uniform exposure to COVID-19 and a higher prevalence of the COVID-19 hospitalization outcome in our cohort. We defined our outcome as a claim for a hospitalization with a positive COVID-19 test between January 1, 2020 and June 26, 2020.

Prescription drug users were identified by string matching from pharmacy claims for any of the generic names associated with the drug candidate. We considered individuals to be drug-exposed when their total supply days covered ≥80% of days between their first drug use date after July 1, 2019 through January 31, 2020. We considered individuals non-drug-exposed if the individual was never prescribed the drug candidates or drugs in the same therapeutic class, between July 1, 2019 and January 31, 2020. We also included one negative control associated with a known COVID-19 confounder, glucose meters, to assess our analysis pipeline's global confounding control. We considered individuals to be exposed when they have one prescription for a glucose meter between July 1, 2019 and January 31, 2020.

*Study covariates*

For each drug of interest, we extracted the following list of covariates for both drug-exposed individuals and non-drug-exposed individuals:
1. Age
2. Gender
3. Self-reported race and ethnicity
4. Area-specified SES index based on member zip code
5. 2019 diagnoses as selected from the top 200 first three-digit ICD-10-CM code, excluding codes beginning with "Z"
6. Pre-existing conditions defined by diagnosis codes in 2019, including conditions used in the Charlson Comorbidity Index and Elixhauser Comorbidity Index
7. Pre-existing primary treatment-related diagnosis
8. Co-used prescription drug defined as claims between July 1, 2019, and January 31, 2020, for the top 20 therapeutic classes
9. Prior hospitalizations in 2019
10. Count of primary care provider visit in 2019
11. Count of unique drugs prescribed
12. Routine screening adherence in 2019, as indicated by completion of a comprehensive metabolic panel, lipid panel, and complete blood count
13. Flu vaccination in 2019 as a proxy of good health behaviors
14. Special Need Plan: (1) institutional, indicating if a member is from a nursing home; (2) dual plan with Medicaid.

*Controlled study without propensity score matching*

We first selected a list of features using a LASSO model with tuned penalty coefficient based on Bayesian information criteria. The complete list of features includes normalized age, sex, primary treatment-related diagnosis, comorbidity index flags, occurrence flags to first three digits of diagnosis codes, adherence flags to co-used drug therapeutic classes, race, state of residence, and normalized SES index. After feature selection, we added normalized age, normalized SES index, and primary treatment-related diagnosis into the feature list to control for these factors. To ensure model convergence, we excluded features with a prevalence of less than one percent of the cohort. We then fit a Cox proportional hazard model to determine the adjusted hazard ratio of the treatment group, considering time to COVID-19 hospitalization, controlling for the list of features selected. We allowed baseline time to vary by individual, setting individual baseline time to be time in our database of first COVID-19 hospitalization for an individual residing in the same state.

*Controlled study with propensity score matching*

For the group of drug-exposed individuals, we applied 1:1 propensity score matching (PSM) to construct a matched group of non-drug exposed individuals. The propensity score was built using logistic regression based on age, sex, primary treatment-related diagnosis, comorbidity index flags, occurrence flags to first three digits of diagnosis codes, adherence flags to co-used drug therapeutic classes, race, state of residence, and SES index. We ran 1:1 PSM with a caliper of 0.25 multiplied by the standard deviation of propensity scores. We assessed PSM performance by calculating the standardized mean difference between drug-exposed and non-exposed groups across the primary treatment related diagnosis. PSM is considered adequate when the standardized mean difference between groups is ≤ 0.10 (Zhang et al., 2019). After PSM we report the unadjusted hazard ratio for the drug-exposed group. In addition, we applied the same procedure of feature selection and similarly fit a Cox proportional hazards model for each drug of interest, between baseline (the state-specific time of first COVID-19 hospitalization) to hospitalization or end of follow-up, to investigate the adjusted hazard ratio of the drug-exposed group. We applied a Benjamini-Hochberg correction with false discovery rate 0.1 to control for multiple hypothesis testing.

*Code availability*

A complete implementation of GSPA, including precomputed embeddings, will be made available as a Python program at https://github.com/henrycousins/gspa.

**RESULTS**

*Overview of GSPA*

Network topology-based gene set analysis methods rely on the principle that aggregate expression changes in gene sets are better resolved by considering expression changes in local gene subnetworks. GSPA extends this principle to a learned latent space that reflects genes' functional similarity through a low-dimensional representation of their contexts in a complete PPI network. GSPA calculates the enrichment of a gene set in a ranked list in an analogous manner to GSEA, leveraging the principle that genes with highly similar embeddings are likely to share functional overlap. Specifically, GSPA computes an enrichment score through the same weighted Kolmogorov-Smirnov statistic used in GSEA, but considering the union of the original gene set and the set of proximal genes meeting a predefined level of embedding similarity to any member of the original gene set.

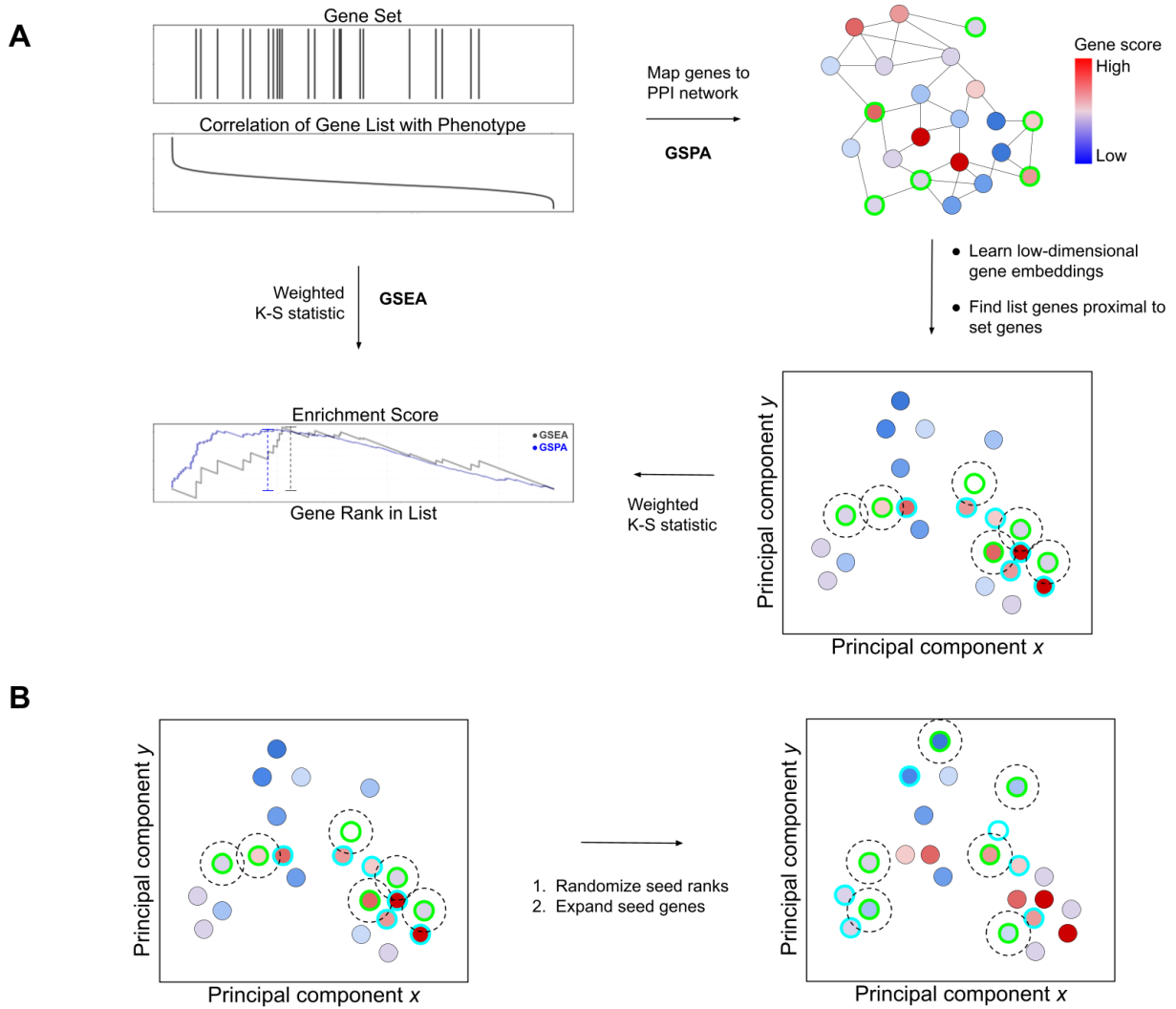

Figure 1. Overview of gene set proximity analysis algorithm
(A) Beginning with a gene set and a list of genes ranked by a continuous score, GSPA maps gene IDs to precomputed embeddings representing the genes' positions within a human PPI network. A set of genes (outlined in cyan) proximal to the original gene set (outlined in green) is determined by adding genes that are within a specified distance, in latent space, from any gene in the original set, even if the dataset lacks a ranking for the original set gene itself. An enrichment statistic is computed as the weighted K-S statistic of the augmented gene set.
(B) For every gene set, a null distribution is determined by performing a fixed number of random permutations of the original gene set (equivalently, constructing a random gene set of the same size as the original), then performing the latent expansion step to incorporate proximal genes.

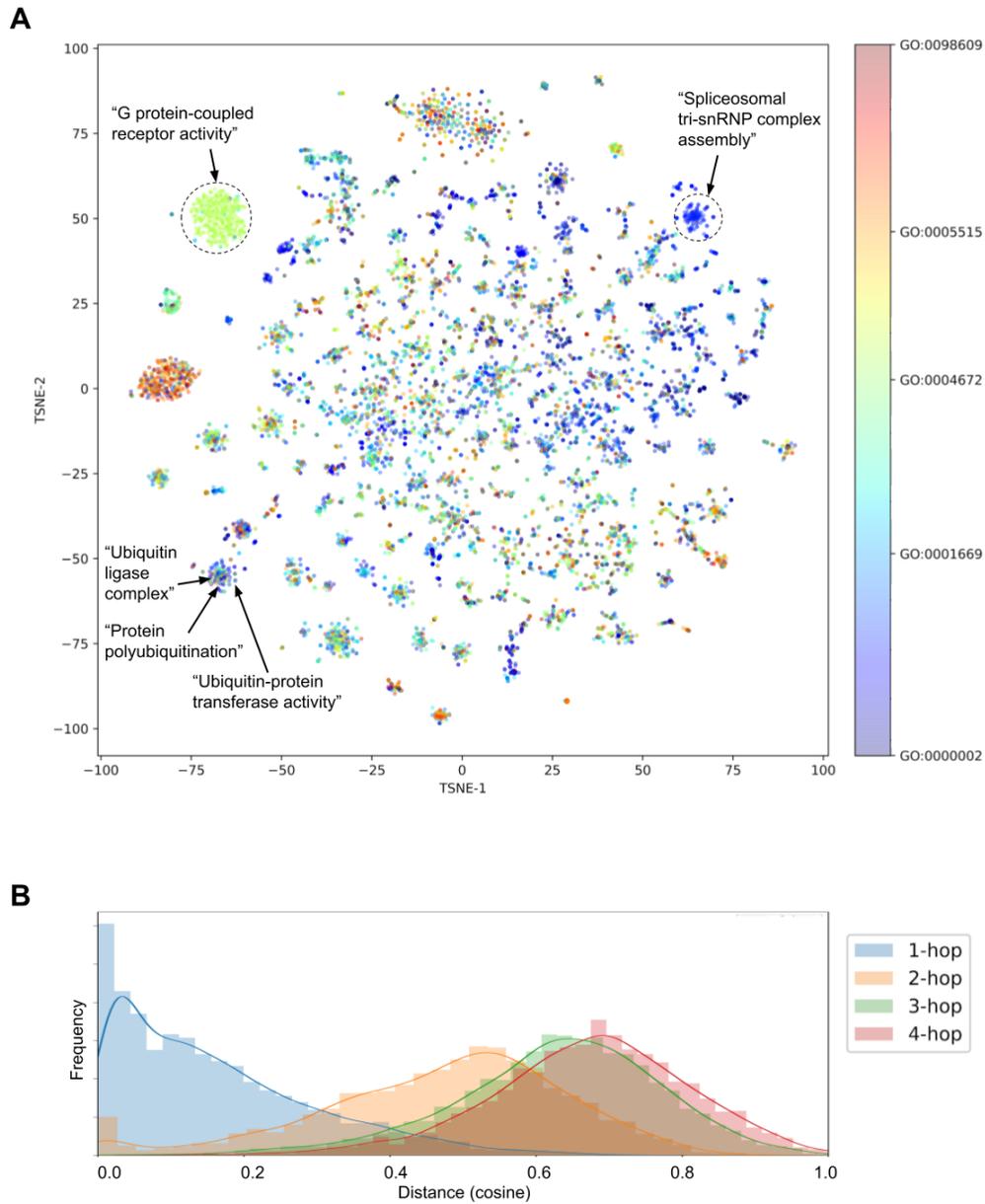

Figure 2. Visualization of unsupervised gene embeddings
(A) T-SNE projections of all gene embeddings after training, colored by GO term association. Functionally associated genes tend to cluster together.
(B) Distribution of embedding similarities for pairs of n-hop neighbors. 1-hop neighbors show the highest degree of similarity (lowest distance), and the distribution of embedding similarities becomes similar for genes separated by more than 3 hops.

*GSPA outperforms GSEA and NGSEA in predicting disease-associated gene sets from expression data*

We evaluated the ability of GSPA, GSEA, and NGSEA to identify known disease-associated gene sets from gene expression datasets for the corresponding diseases. For this analysis, we used the gold-standard GEO2KEGG compendium, containing 42 gene expression datasets matched to 19 different diseases. 15 of these diseases corresponding to 33 gene expression datasets were matched with a set of known disease-associated KEGG pathways previously defined by Geistlinger et al. (2021) for use in evaluating gene set analysis methods. For each expression dataset, we ranked 343 KEGG pathways using GSPA, GSEA, and NGSEA and measured each method's ability to identify known disease-associated pathways by area under the precision-recall curve (AUPRC).

We observed significantly higher predictive ability with GSPA compared to both GSEA and NGSEA ($p = 1.327e-4$ and $p = 0.00458$, respectively). GSPA outperformed GSEA on 25 of the 33 tested datasets (75.8%) and NGSEA on 21 of 33 (63.6%) datasets. GSPA scored highly on datasets for a variety of diseases, including solid and liquid cancers, cardiomyopathy, and Huntington's disease, although performance was relatively reduced in Alzheimer's and Parkinson's datasets.

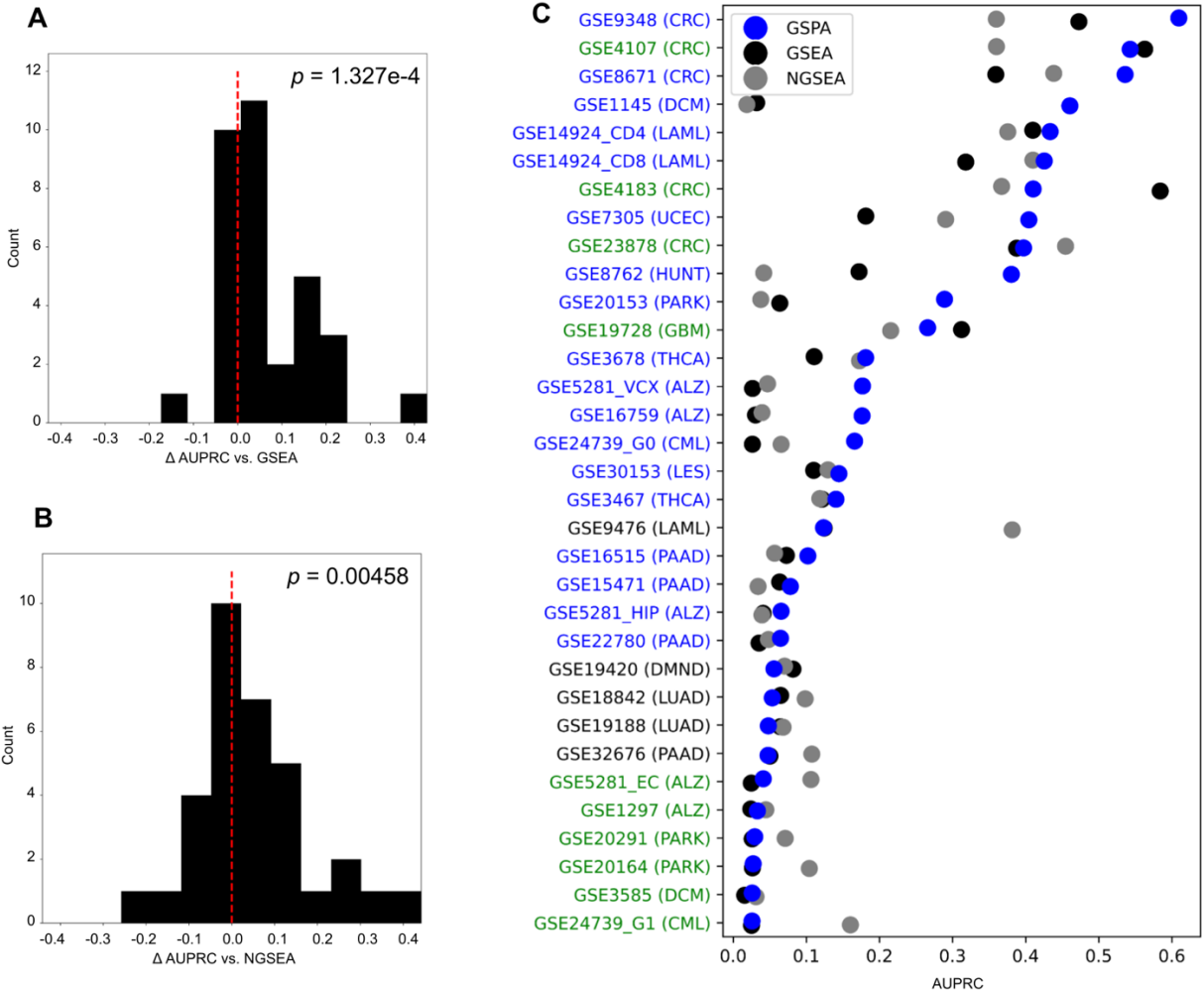

Figure 3. Ability to retrieve known disease-associated gene sets
(A) Difference in AUPRC between GSPA and GSEA for identification of literature-derived, disease-associated gene sets in disease-specific expression datasets. Density to the right of the red line indicates better performance from GSPA.
(B) Comparison of GSEA and NGSEA in the same format as (A).
(C) Performance of GSPA (blue), GSEA (black), and NGSEA (gray) on all individual datasets. Blue text indicates where GSPA performed best and green text indicates where GSPA performed second-best.

*GSPA outperforms GSEA and NGSEA in reproducing rankings for semantically similar gene sets*

We next assessed the ability of GSPA to return semantically consistent results, as traditional gene set analysis methods are sensitive to the specific member-wise composition of a gene set, which is often arbitrary with respect to the gene set's intended meaning. We used an automated procedure to define a set of 54 semantically similar gene-set pairs, such as "hsa00100_Steroid_biosynthesis" and "hsa00140_Steroid_hormone_ biosynthesis", and measured the pairwise correlation between enrichment rankings for each method. We observed significantly stronger correlations with GSPA than with either GSEA or NGSEA (p = 0.00113 and p = 4.97e-4, respectively). GSPA outperformed GSEA on 33 of the 42 tested datasets (78.6%) and NGSEA on 31 of 42 (73.4%).

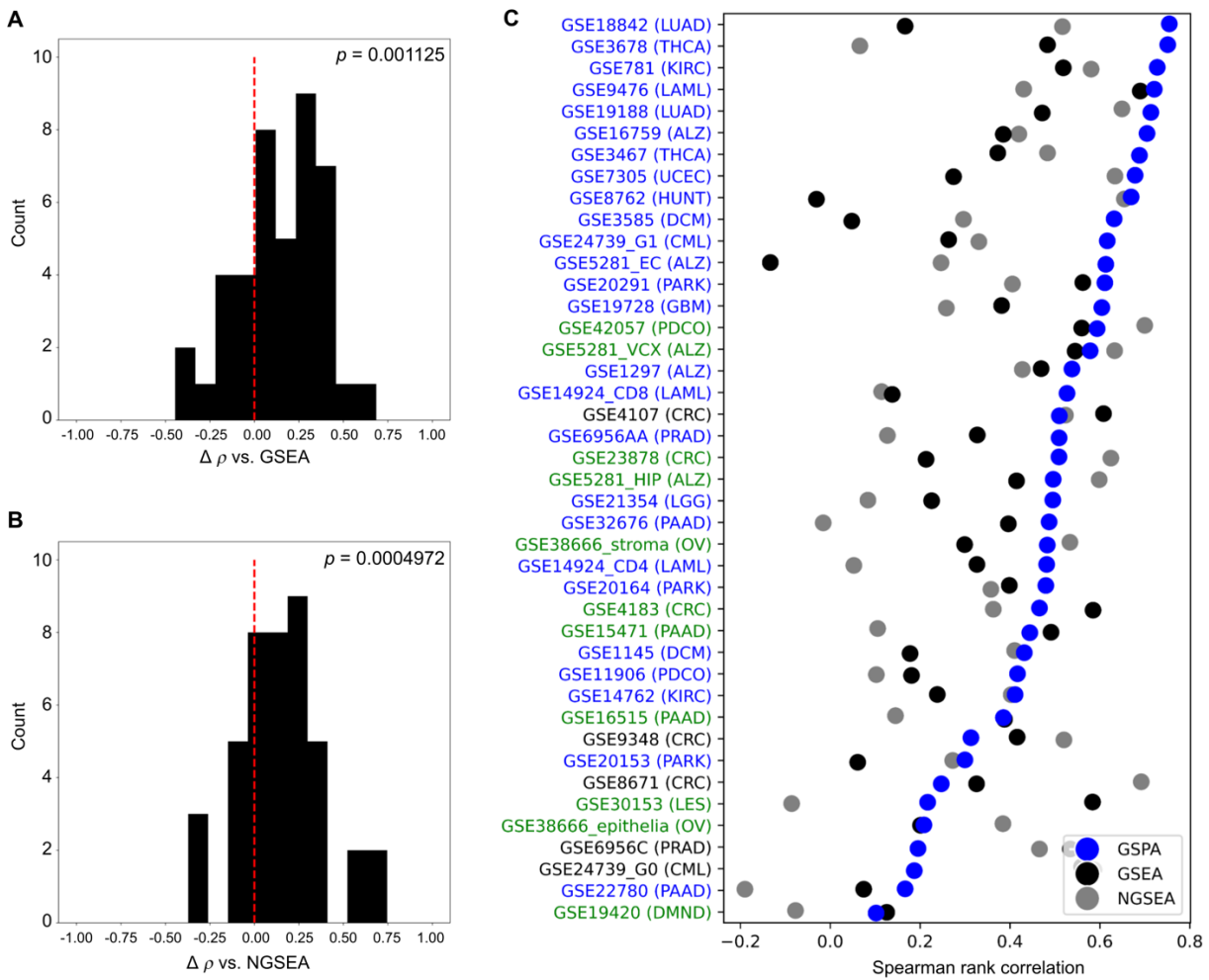

Figure 4. Reproducibility among semantically similar gene sets
(A) Difference between GSPA and GSEA of Spearman correlation between rankings of matched gene sets representing the same pathway. Density to the right of the red line indicates better performance from GSPA.
(B) Comparison of GSPA and NGSEA in the same format as (A).
(C) Performance of GSPA (blue), GSEA (black), and NGSEA (gray) on all individual datasets. Blue text indicates where GSPA performed best and green text indicates where GSPA performed second-best.

*GSPA enables prediction of pharmacologic modulators of SARS-CoV-2 viral entry*

Gene set analysis methods are valuable tools for drug discovery and repurposing, as they provide a means of associating specific pharmacologic agents with disease phenotypes by means of aggregate expression changes in gene ensembles targeted by the same compound.

We investigated whether GSPA could be useful for drug repurposing efforts by measuring its ability to retrieve known disease-drug associations. We first obtained gene sets representing known targets of FDA-approved drugs from the Drug Signature Database (DSigDB) and compiled drug-disease associations from the Comparative Toxicogenomics Database. For each of our evaluation datasets, we measured the ability of GSPA and GSEA to identify gene sets representing drugs with known associations with the corresponding disease. GSPA demonstrated clear benefit on this task, with a significant ($p = 0.0128$) improvement in AUPRC compared to GSEA.

We next investigated whether GSPA could identify novel drug repurposing opportunities in COVID-19. Given the growing amount of literature supporting a multifactorial viral entry mechanism influenced by many host genes, we focused on identifying approved drugs able to modulate viral entry into host cells. We obtained three datasets representing genome-wide CRISPR knock-out screens ranking host genes by their associated effect on SARS-CoV-2 viral entry. We ran GSPA with drug-target gene sets separately on each dataset, generating a ranking of 468 FDA-approved drugs. Notably, all three datasets revealed an enrichment of drug classes with well documented COVID-19 associations, including corticosteroids and atypical antipsychotics (Nemani *et al.*, 2021; Mangin and Howard, 2021; Gordon *et al.*, 2020). For follow-up analysis, we restricted our search to highly prescribed, systemic drugs (within the top 100 in 2020 in the US), of which only four drugs had a false discovery rate below 0.5 in any dataset: the benzodiazepines clonazepam and lorazepam, gabapentin, and metformin. Furthermore, we noted that gabapentin, clonazepam, and lorazepam were associated with positive normalized enrichment scores, while NES values for metformin were consistently negative. This implies that metformin would be predicted to exert an opposite effect on viral entry in comparison to gabapentin, clonazepam, and lorazepam.

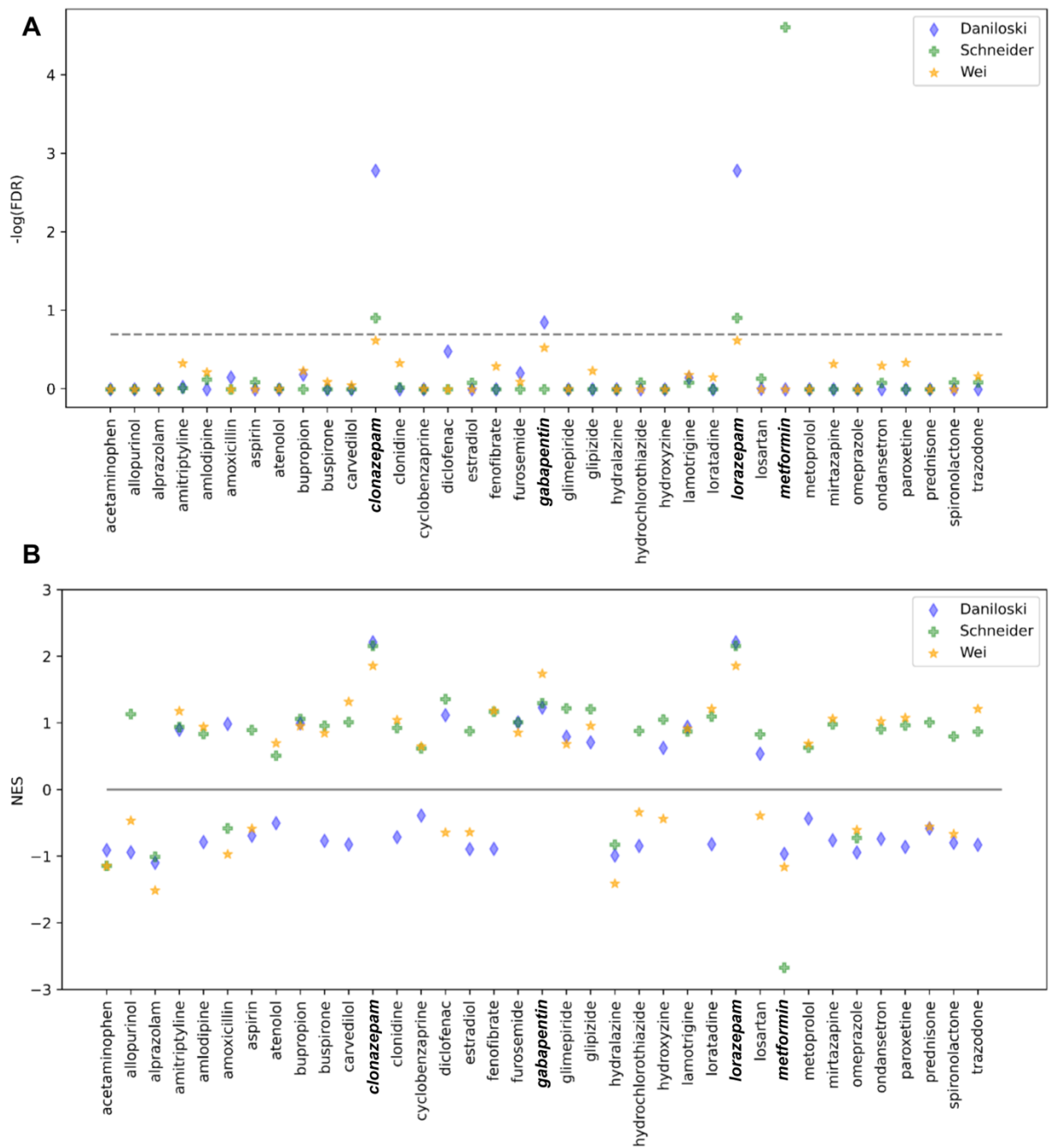

Figure 5. Analysis of SARS-CoV-2 entry datasets using DSigDB gene sets
(A) GSPA FDR scores for commonly prescribed, FDA-approved drugs on three datasets of gene essentiality for SARS-CoV-2 viral entry. Dashed line indicates an FDR threshold of 0.5.
(B) Same as (A) but showing NES. Positive and negative NES indicate opposite predicted effects.

*Retrospective analysis of health insurance claims supports a role for metformin and gabapentin in modulating SARS-CoV-2 viral entry*

To investigate whether prediction of a modulatory effect on SARS-CoV-2 viral entry correlates with a drug's clinical effect, we performed a retrospective analysis of claims data from a large US health insurance provider, examining associations between common prescriptions and COVID-19 hospitalization rates. We reviewed claims from 7.8 million Medicare Advantage Part D (MAPD) members for compatibility with regional and temporal inclusion criteria. The final dataset comprised claims for 234,524 MAPD-insured residents of New York, New Jersey, and Connecticut with at least 11 months of enrollment between January and December 2019 and at least one month of enrollment during 2020, with at least one pharmacy prescription claim. Among these individuals, 2,828 (1.21%) had claims indicating COVID-19 hospitalization during the observation window.

For each of the four candidate drugs, we measured the association between drug use and hospitalization due to COVID-19 both with and without 1:1 propensity score matching (PSM). After controlling for multiple hypothesis testing using a Benjamini-Hochberg correction, neither clonazepam nor lorazepam was associated with a significant hazard ratio for COVID-19 hospitalization. However, gabapentin use was associated with a significantly elevated risk of COVID-19 hospitalization, with an adjusted hazard ratio of 1.211 (p = 0.0020) before and 1.189 (p = 0.034) after applying PSM. Furthermore, after applying PSM, metformin use was associated with a reduction in COVID-19 hospitalization risk, with an adjusted hazard ratio of 0.834 (p = 0.013).

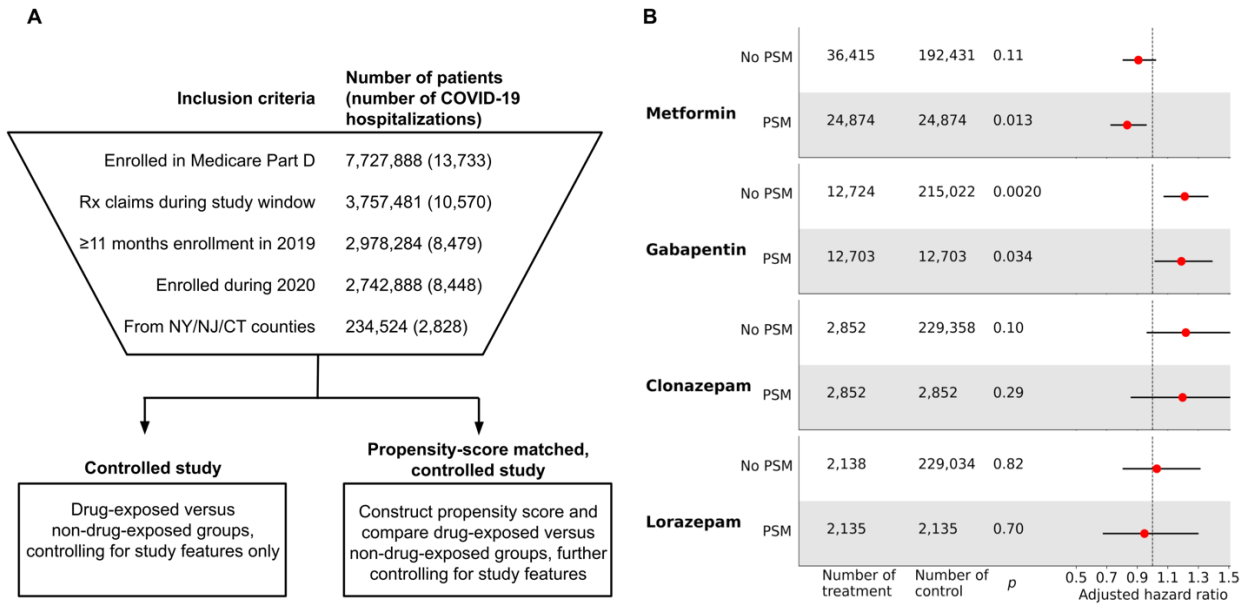

Figure 6. Retrospective clinical analysis of drug effect on COVID-19 hospitalization
(A) Overview of analysis procedure. From 7.7 million database patients, 234,524 met criteria for inclusion. Two comparisons were performed, comparing COVID-19 hospitalization rates among drug-exposed versus non-drug-exposed patients, controlling for covariates, with and without propensity score matching.
(B) Results for drug hits identified by GSPA. Metformin shows a significant negative association with hospitalization, while gabapentin shows a significant positive association with hospitalization.

## DISCUSSION

In this work, we present a novel algorithm, GSPA, based on unsupervised graph learning that extends traditional gene set enrichment analysis to a latent feature space reflecting complex interactions among protein-coding genes. While some existing methods for gene set analysis use a priori knowledge of pathway structures in the form of PPI networks, most such methods work by measuring distance between two distinct sets of genes, requiring arbitrary thresholding of experimental measurements to establish a set of differentially expressed genes. Of the few network-augmented methods that consider continuous experimental scores, all use explicit representations of network structure, requiring either consideration of only the immediate neighborhood of a gene of interest (Han *et al.*, 2019) or long runtimes (Nadeau *et al.*, 2021). GSPA derives from the theory that unsupervised graph embeddings, popular in social network analysis, provide lightweight and expressive representations of PPI network topology. This notion provides a demonstrated basis for identifying functional gene set members (Wang *et al.*, 2020). However, simply applying GSEA methods to the enhanced gene sets resulting from network-based expansion methods results in nonspecific significance estimates, as randomly derived null distributions fail to account for presumptive associations between expression changes between PPI network neighbors (Goeman and Bühlmann, 2007). The null distribution calculated by GSPA implicitly accounts for known associations, providing more specific significance estimates than do other approaches.

To our knowledge, GSPA is the first gene set analysis method to incorporate implicit representations of pathway features, which provides several advantages in comparison to other network-augmented gene set analysis methods. First, learned embeddings capture both local and global genetic features that would not be apparent using simpler gene-level metrics such as shortest path length to a gene set. They are also modular, allowing any user-defined vector representation of gene similarity to be substituted, and require significantly less computational expense than do operations on unreduced networks. Furthermore, the use of embedding similarity to determine proximal genes enables a more conservative and biologically reasonable null hypothesis that accounts for known gene-gene associations, which is not possible in most gene set analysis methods for ranked lists. Finally, the GSPA algorithm accommodates arbitrary user-defined gene similarity thresholds, reducing exactly to the classical GSEA algorithm as this parameter becomes highly stringent.

We apply GSPA to a variety of gene set analysis tasks, showing that it provides improved performance with respect to detection of disease-associated pathways in gene expression datasets, compared both to GSEA and to NGSEA, a state-of-the-art network-augmented approach. GSPA achieved high performance across a variety of disease types, including solid cancers, leukemias, neurodegenerative diseases, and chronic cardiopulmonary diseases. We observe that GSPA tends to perform worst on neurodegenerative diseases, which was also observed for NGSEA. We hypothesize that this phenomenon may derive from relative underrepresentation of these conditions in literature-based PPI networks, in comparison to cancers (Szklarczyk *et al.*, 2017). For specialized applications in specific diseases, fine-tuning gene embeddings based on disease-specific literature may improve performance in the future (Zhao *et al.*, 2021; Ietswaart *et al.*, 2021).

We also demonstrate that GSPA improves reproducibility of enrichment statistics for gene sets with a shared semantic meaning. A common problem in traditional gene set analysis methods is their sensitivity to small changes in the definition of knowledge-based gene sets. For instance, for the diabetes mellitus dataset assessed in the original GSEA report, the MSigDB C2 gene set "p38mapkPathway" was ranked 20th in enrichment, while "ST_p38_MAPK_Pathway" was ranked 117th out of 318 (Subramanian *et al.*, 2005). Such variance reflects discrepancies between the specific definition of a gene set and the underlying pathway that it represents semantically, impairing both interpretability and reproducibility of enrichment analyses. By representing gene sets implicitly in a latent feature space, GSPA reduces the sensitivity of the enrichment test to these discrepancies.

Finally, we use GSPA to make novel predictions of drug associations with SARS-CoV-2 infection. We apply the algorithm on three datasets measuring genome-wide gene essentiality for viral entry in lung and epithelial host cells using gene sets representing known targets of FDA-approved drugs. This yielded four commonly prescribed drugs with a predicted modulatory effect on SARS-CoV-2 entry: metformin, gabapentin, clonazepam, and lorazepam. While gabapentin, clonazepam, and lorazepam targets were significantly enriched toward the top of the datasets, metformin targets were significantly enriched toward the bottom, suggesting an opposing effect of metformin in comparison to the other drugs.

We validated the drug predictions through retrospective analysis of COVID-19 hospitalization rates in propensity-score-matched subjects with or without exposure to each drug, observing statistically significant hazard ratios associated with both gabapentin and metformin. Specifically, we observed that gabapentin was associated with increased risk of COVID-19 hospitalization, while metformin was associated with decreased risk, consistent with the opposing effects predicted by GSPA. Several retrospective studies have demonstrated an association between metformin use and COVID-19 hospitalization and mortality, supporting the generalizability of our clinical findings (Lalau *et al.*, 2021; Luo *et al.*, 2020; Crouse *et al.*, 2021). However, the potential role of gabapentin as a risk factor for viral entry has not been investigated outside of our study. Further experimental and clinical investigations are necessary to clarify the effect size and mechanism of action of gabapentin in COVID-19 patients.